# Conversation Analysis and the User Experience

Allison Woodruff and Paul M. Aoki
Palo Alto Research Center
3333 Coyote Hill Road
Palo Alto, CA 94304-1314  USA
woodruff@acm.org, aoki@acm.org

**ABSTRACT**
We provide two case studies in the application of ideas drawn from conversation analysis to the design of technologies that enhance the experience of human conversation. We first present a case study of the design of an electronic guidebook, focusing on how conversation analytic principles played a role in the design process. We then discuss how the guidebook project has inspired our continuing work in social, mobile audio spaces.  In particular, we describe some as yet unrealized concepts for adaptive audio spaces.

**INTRODUCTION**
The sociological discipline of conversation analysis [9] (hereafter *CA*) has long held a significant position in user experience design.  The idea that ethnomethodology, and CA in particular, can have direct application to design is widely credited to Suchman [12], who not only proposed their methodological use in the study of human-machine interaction but also observed that an awareness of human social practices of (e.g.) conversational repair can itself serve as a resource for design.  Since then – even leaving aside studies of technology use that apply conversation analytic methods – a number of attempts have been made to apply CA to HCI in a very direct way; we briefly discuss some of these in [15].

Given this history, which is nearly as long as that of the SIGCHI conference itself, it may seem odd to speak of CA as anything but a traditional methodological source for experience design.   "Tradition" in the sense of history, however, does not imply "traditional" in the more common sense of customary or characteristic use.  Most popular design methods, such as contextual design, are broadly applicable and can be learned from courses or textbooks. By contrast, CA focuses on human-human interaction, and a professional level of proficiency in CA methods is best attained through long practical apprenticeship.  As a result, researchers often find it difficult to apply CA to design in ways that are both productive and consistent with its sociological outlook.

Nevertheless, CA can be very helpful in system design, most clearly for systems that involve spoken language. We base this claim on our own experience – for several years, we have been drawing on CA to inform the design of computing technologies that are intended to facilitate aspects of *human-human* interaction.  In a previous paper [15], we attempted to illustrate these uses and provide "how to" instructions for incorporating a trained conversation analyst into the iterative design process.  We continue that discussion here, showing how CA has led us into new research areas.

Our story consists of two interrelated narratives.  After a brief description of CA, we discuss our project on the design of an electronic guidebook for historic houses that facilitates social interaction between visitors.  This is mature research, largely conducted during 1999-2002; the discussion demonstrates the use of CA to identify conversational structures that are important in facilitating users' social goals.  We then turn to a description of an ongoing project on the design of mobile audio communication systems.  The idea is to provide communication technologies that actively facilitate social interaction by monitoring spoken conversation in a mediated communication channel, recognizing the presence of specific conversational structures, and then changing specific properties of the communication channel to support the social goals implicit in the use of these structures.  This is research that has resulted in some early prototypes but is still very much in-progress.

**CONVERSATION ANALYSIS**
Conversation analysis, the most visible and influential form of ethnomethodological research, is concerned with describing the methods by which the members of a culture engage in social interaction [9].  A key goal of CA is to examine social interaction to reveal organized practices or patterns of actions, under the fundamental assumption that interaction is structurally organized.  Social actions include talk, gesture, and use of objects.  While ethnomethodology and CA share this concern for how actions are organized, the goal of CA is to describe both how sequences of action are organized and situated in a particular instance of activity, as well as to abstract features that *generalize* across a collection of similar instances.

A conversation analytic research program involves analyzing a collection of interactive encounters.  The analysis is twofold.  First, the analyst makes a moment-by-moment, turn-by-turn transcript of the actions in each encounter.  Second, the analyst examines these encounters individually and then comparatively to reveal a practice's generalizable orderliness.

To make the discussion more concrete, consider the procedure we used in our own studies of electronic guidebook use. As we discuss further below, we use CA to describe visitors' systematic practices as they use an electronic guidebook to tour a historic house with a companion. To identify these systematic practices, we examine in detail the data collected during selected visits. Specifically, for each visit, we create a video that includes the audio and video recordings of the visitors, as well as audio of the descriptions and video of the screens of each visitor's electronic guidebook. The resulting data are transcribed and analyzed. Both of these steps require careful, repeated viewing of each video.

**COMPLETED WORK: ELECTRONIC GUIDEBOOKS**
In this section, we review the salient aspects of some work we have done on handheld audio guidebooks for historic houses. This information has been reported elsewhere in more detail [2,14,15] but it forms an important backdrop for the section that follows.

Visitors often go to cultural heritage locations, such as museums, with companions. Many seek what has sometimes been called a "learning-oriented" experience [6]. To facilitate learning, institutions typically present information through guidebooks and prerecorded audio guides as well as through labeled exhibits and docent-led tours. However, sharing the experience with companions is often a higher priority than learning, particularly for infrequent visitors [7]. Unfortunately, typical presentation methods interfere with the interaction among visitors. For example, visitors frequently complain that audio tours with headphones isolate them from their companions, and visitors have few opportunities to interact effectively with each other while docents "lecture" to them.

Our project had the goal of designing an electronic guidebook that would facilitate rather than hinder social interaction. To achieve this goal, our project followed an iterative design process. We designed and implemented two major prototypes (as well as a number of more incremental prototypes). We observed visitors using the prototypes during self-guided tours of a historic house and we conducted semi-structured interviews about their impressions of the guidebook. Visitor actions during the tour were captured using audio recording, video recording, and device logging. We used CA to analyze people's interactions with each other and their use of the guidebook. Based on our findings, we formulated new design principles and revised the prototypes. In the remainder of this section, we discuss each prototype and relevant findings associated with it.

The first prototype was designed to provide a range of options for information presentation and sharing. For example, it included options for textual descriptions, audio descriptions delivered through speakers played into open air, and audio descriptions played through headphones. Analysis of the use of this prototype revealed that visitors had a strong preference for speaker audio since it afforded a shared listening experience for companions. Further, CA revealed that when visitors had a shared listening experience, they oriented to the guidebook as though it was a human participant [14]. This was achieved through careful audio design (e.g., using audio clips that did not exceed expected human turn lengths; scripting audio content in ways that a human speaker would "design" a conversational turn; ensuring that listeners could hear audio clips simultaneously). Visitors structured their conversations around the guidebook's audio, creating a place for it in their social interaction, e.g., visitors made a place for the guidebook to take turns in the conversation. More specifically, CA demonstrated that visitors oriented to the guidebook descriptions as though they were stories, following discourse patterns that have previously been observed with human storytellers [8]. This was desirable since it meant that the flow of the visit could take the form of an ongoing conversation between visitors into which the guidebook content could be fitted, as opposed to a series of long lectures from the guidebook.

To minimize "audio clutter" when a large number of visitors wish to play descriptions simultaneously, we designed a second prototype with a technological mechanism that allows visitors to have a shared listening experience with headphones [2]. Specifically, devices are paired and communicate via a wireless network. Each visitor in a pair always hears the content they select themselves, and additionally, each visitor has a volume control for determining how loudly they hear content from their companion's guidebook.

CA studies of the use of this second prototype indicated that the shared listening experience was preserved and that the visitors continued to orient to the guidebook as a storyteller. Further, the analysis revealed interesting patterns of engagement. In general, when people are gathered together and involved in an activity, conversational interaction may occur, then lapse, then occur again. After a lapse, people *re-engage* turn-by-turn talk; alternatively, when people suspend turn-taking and *dis-engage* turn-by-turn talk, a lapse occurs. To accomplish states of re-engagement and dis-engagement, people draw upon a wide range of verbal and non-verbal communicative resources as well as features of the activity in which they are involved. Our studies showed that the guidebook was a resource for sustaining engagement and for re-engaging conversational partners. The second prototype was significantly more effective for this purpose compared to the first because of subtle changes in the design, such as the increased ease of listening with headphone versus speaker audio. The fact that subtle changes had such a dramatic impact on conversational structures led us to a new area of research.

**CURRENT WORK: SOCIAL, MOBILE AUDIO**

In our current research, we have taken as a design challenge the problem of creating *remote* audio communication technologies that actively facilitate remote conversations that are "more like" co-present conversations. Specifically, we hypothesize that explicit technological facilitation of conversational structures like those arising in co-present conversation can enhance the experience of casual, spontaneous, social conversation using remote audio.

Our project's CA work on electronic guidebooks led immediately to this area of inquiry. Recall that the key phenomena from the first guidebook study (using the speaker audio prototype) related to the sequential organization of storytelling, but that the most striking aspects of the subsequent studies (using the remote eavesdropping prototype) related to the structures employed by co-present speakers as they engage, dis-engage, and re-engage in conversation. This focus on engagement and on spontaneous, "on-again, off-again" talk led us to consider how we might make remote audio communication more like the "continuing state of incipient talk" [11] afforded by co-presence.

At present, most remote audio communication consists of telephone conversation, which differs from co-present conversation in that it exhibits relatively heavyweight *openings* and *closings* [11]. However, previous research on the desktop use of "always on" audio communication environments, known as *audio spaces* [1], suggests that when people remain connected by an open audio communication channel over a period of time, structures arise that resemble re-engaging and dis-engaging talk. Can we do better than what is basically an open conference call?

**Facilitating Social Multi-party Conversation**

One probe in this direction is a system that we have designed to facilitate lightweight group discussion within an audio space [3]. From the literature and our own design ethnography [13], we observed that the highly dynamic structure of *social multi-party conversation* was poorly served by existing audio communication systems. (Consider how difficult it is to have a spontaneous conversational experience, like that around a dinner-table or at a cocktail party, in a telephone conference call.) A major reason for this is that existing systems have little support for *schisming* – the transformation of one conversational floor into two simultaneous conversational floors, which is common in such co-present conversations. Our audio space system includes a machine learning component that analyzes participant turn-taking behavior to identify conversational floors as they emerge, noting which participants are in which floor. The system dynamically modifies the audio delivered to each participant to enhance the intelligibility of the participants with whom they are currently conversing and to reduce the salience of the participants with whom they are not currently conversing. Each participant therefore receives a customized mix of all floors, tailored to their current conversational status.

The system applies some direct corollaries of CA research to decide who is talking to whom. The organization of taking turns at talk is fundamental to conversation. One of the ways in which turn-taking organization operates is by specifying opportunities for speaker change at *turn-constructional units* (TCUs) from which turns at talk are composed [10]. This enables listeners to monitor and project the completion of others' TCUs in order to time the initiation of their own turns properly. Completion of a TCU is often accompanied by a *pause* in speech, making a *transition-relevance place* (TRP) where speaker change may occur. Multi-party conversations may consist of a single floor in which participants orient to each others' turn-taking behavior as just described. However, in casual multi-party conversation, a given floor frequently schisms into multiple floors and multiple floors frequently merge [5]. When two simultaneous conversational floors are on-going, participants in one do not orient to the turn-taking organization of the other. In CA terms, this implies that TCUs of people in the same floor will have different statistical distributions of pause and overlap duration relative to TCUs of people in different floors; we approximate this by measuring the pause and overlap duration of speakers' utterances and comparing them to pre-learned statistical models.

We built an proof-of-concept prototype of this system and performed a preliminary evaluation [3]. When the system operates properly, the resulting experience is much like that in co-present conversation – you can easily follow the speech of people with whom you are conversing, and others can be understood with effort. When the membership of a floor changes, mutual intelligibility adapts accordingly. We continue to work to improve the system's effectiveness.

**Smoothing Conversational Engagement**

We are currently working on a second design concept that is intended to facilitate remote conversation. Unlike the previously described system, this has not yet been realized as a complete implementation.

The findings of our design ethnography also suggested that speakers in various degrees of conversational engagement exhibit different conversational *styles* that can be characterized by the kinds of gaps that can occur between turns at talk. Further, as engagement varies, speakers may prefer certain types of communication media over others. (Consider the fact that instant messaging sometimes seems more suitable than the telephone, and vice versa.) This behavior of changing communication devices or applications is known as *media-switching*. Since re-engagement and dis-engagement can be highly dynamic, it is highly desirable to support users in their moment-by-moment changes of conversation style with maximal fluidity, i.e., without requiring them to switch devices or applications. In contrast to planned and negotiated media-switching, we propose that a more spontaneous experience is afforded if the channel itself adapts appropriately to users' conversation, thereby supporting what we call *style-switching*. (These points are elaborated further in [13].)

We suggest that technological means can be used to adapt a channel to participants' conversational needs [13]. As an example, consider a system that monitors participants in an ongoing conversation and automatically adapts properties of the channel – properties that have, in the past, been fixed for a given type of channel, such as half-duplex vs. full-duplex – in response to observed characteristics associated with different conversation styles. Such characteristics can be of the individual participants (e.g., their observable emotional state), or of their interaction (e.g., their turn-taking engagement with other participants). For example, imagine two participants in a push-to-talk session, each responding slowly because they are both busy with other tasks. Now suppose that a new topic of conversation is raised and both participants become highly interested. The system may detect that the participants are showing strong signs of interest (e.g., their voices have acoustic properties correlated with interest) and that they are showing signs of increased conversational engagement (e.g., they begin to respond much more rapidly than before). In response, the system shifts the channel to an open-microphone, full-duplex mode, playing a short tone to indicate that push-to-talk will no longer required. Later, when the demonstrated level of engagement dies down (e.g., by a sustained pattern of lapses between turns), the system shifts the channel back to push-to-talk.

**CONCLUSIONS**

We have provided some brief examples of how ideas drawn from CA can be used to enhance the experience afforded by technologies that are designed to play a role in human conversation. In particular, we have shown how we have applied an understanding of various types of sequential organization known to the CA community (e.g., storytelling, dis-engaging and re-engaging talk in activity, schisms) to the design of systems that facilitate specific aspects of human conversation.

We note in closing that it should be remembered that CA is about human-human interaction. We do not try to design computer systems that purport to interact with humans according to the "rules" of CA (a claim that draws loud complaint from the CA community [4]). Rather, we design systems to operate in a manner that reflects *human* practices that are likely to arise in particular situations. Whether this is a passive compatibility (as in the case of the electronic guidebook) or an active behavior (as in the case of the adaptive audio space), we have found the CA perspective very helpful in making the experience of mediated conversation more natural.

**ACKNOWLEDGMENTS**
The research described here resulted from past and on-going collaborations. In addition to the team conversation analyst, Peggy Szymanski, the team has included Jim Thornton, Beki Grinter and several summer interns. Any errors and misrepresentations, however, are ours alone.